\documentclass[twocolumn,showpacs,preprintnumbers,amsmath,amssymb,prl]{revtex4}
\usepackage{graphicx}
\usepackage{dcolumn}
\usepackage{bm}
\begin{document}
\title{The  laser polarization as control parameter in the  pattern formation}
\author{S.Varlamov, M.Bestehorn, O.Varlamova and J.Reif}
\affiliation{Institut f\"ur Physik, BTU Cottbus, PBox 101344,
         03013 Cottbus, Germany}
\date{\today}

\begin{abstract}
The recently observed dependence of the periodic surface structures
on the light polarization in the laser induced pattern formation
is analyzed within a model 
where the polarization induces significant deviation the spatial
distribution of the energy deposited by the photon from isotropic
energy distribution. We argue that 
the laser polarization breaks 
the rotation symmetry on the surface and is responsible for the correlation of 
the surface structures with the degree and the direction of polarization.
Moreover it is shown that the polarization induces the appearence of novel features
of the surface morphology and time evolution, which could be directly tested experimentally.
\end{abstract}
\maketitle

The phenomenon of ripple formation on the surfaces of eroded materials is
of particular interests due to the large area of application in physical,
chemical and material sciences. 
The development of periodically modulated structures as result of
sputtering, the removal of atoms from the surface of solids through
the impact of energetic particles (photons or ions), 
was discovered and studied experimentally nearly three decades ago. 
The first widely accepted theoretical approach \cite{EMMONY,SIPE} describing
the periodic surface structures induced by laser radiation suggested that
the ripples are the result of interference of incoming laser beam with some form
of a surface-scattered electromagnetic wave.
In general, this theory
was successeful in discription of uniformly distributed patterns
with the periodicity dependent from the wavelength of laser radiation
and from the angle of incidence.

The development of ultra-short pulses laser technologies and application
of femtosecond laser radiation in surface sputtering
\cite{HENYK} has shown that newly rediscovered laser induced periodic
surface structures (LIPSS), with lateral periods a few times smaller than 
the wavelength of the incident light, can not been described 
in the framework of the conventional LIPSS theory \cite{BONSE}.
Thus, it turns out that the ripple structure has
non-trivial surface morphology sharing many similarities with aeolian
sand dunes \cite{MISBAH}, with periodicity independent from
laser wavelength and the angle of incidence, but correlated with
the local intensity of the laser beam that gives the strong support
to the nonlinear self-organized mechanism of formation this structures.
In this light the understanding of the puzzling dependence
the ripple orientation from the laser polarisation \cite{HENYK,VARLAMOVA}
where the resulting ripple orientation depends on the direction
of the vector of electromagnetic field,
become very important and interesting problem.

A revival of interest in the ion induced ripple formation 
has been caused by successful theoretical prediction \cite{BRADLEY} 
the ripple wavelength and orientation in agreement with experimental observations.
Further development the nonlinear theory \cite{CUERNO,MAKEEV,KIM,CASTRO} 
allowed the proper description
the time evolution of the surface morphology under ion-bombardment,
ripple stabilization, wavelength dependence with ion energy or flux
and production of dot structures as a function of bombardment conditions.

In this ${\it Letter}$ we develop a continuum theory of erosion by
polarized laser radiation. We exploit connections with ripple
formation by ion-beam sputtering and extend this model with
inclusion of laser polarization, leading to polarization dependence of 
coefficients in nonlinear equation of the Kuramoto-Sivashinsky type. 
Our results suggests that the laser polarization can be very
important control parameter.

Following to the Sigmund's theory of sputtering \citep{SIGMUND} we consider
the normal erosion velocity at the surface
\begin{equation}
V \cong \Lambda\int_{\mathcal{R}} d{\bf r} \Psi({\bf r}) E({\bf r}) ,
\label{erosion:rate}
\end{equation}
where $\Psi({\bf r})$ is the local correction to the uniform flux
due to variation of the local slopes, $\Lambda$ is a material
constant and $E({\bf r})$ is the average energy deposited at point O due 
to the scattering of the photon flux at P
\begin{equation}
E({\bf r'})=\frac{\epsilon}{(2\pi)^{3/2}\alpha\beta\gamma}
\exp\left[ -\frac{x'^{2}}{2\alpha^2}
- \frac{y'^2}{2\beta^2} - \frac{z'^2}{2\gamma^2} \right].
\label{gauss}
\end{equation}
$\epsilon$ is the total energy carried by the photon flux and $\alpha$, $\beta$ and 
$\gamma$ are the Gaussian distribution widths along x',y' and z' axes,
respectively, in the reference frame of the incoming beam. 
According to Sigmund's theory  the energy distribution widths 
$\alpha,\beta,\gamma$ along $x',y',z'$ axis, respectively, scaled as
$\alpha=\beta \gtrsim \gamma \backsim $ a - average depth of energy deposition.
Obviously, the Gaussian form is not universal  and the condition $\alpha=\beta$ is not
a general case, in particular in the case  of light interaction with a matter. 
Whereas the deviations from the Gaussian only slightly modified the pattern formation
process \cite{FEIX}, the asymmetry of  energy distribution along surface plain has not been studied yet.
One can expect, that the space anisotropy of the energy deposition can
be effectively described by the anisotropy of the distribution widths $\alpha$ and $\beta$.
Therefore, we except the Gaussian form of energy deposition and consider the laser radiation
which penetrates the bulk of the material and stops at some point, where its energy
spread out to the neighboring sites (see Fig.1). 
In order to introduce the polarization in our calculations we choose the vector ${\bf E}$
of electromagnetic field to lie in the x'-z' plane in the reference frame of the incoming beam.
Following to \citep{BRADLEY,CUERNO} we perform the calculation of erosion rate
in the local coordinate system (X,Y,Z) shown in Fig.1.
\begin{figure}[tbh]
\includegraphics[width=8cm,clip=true]{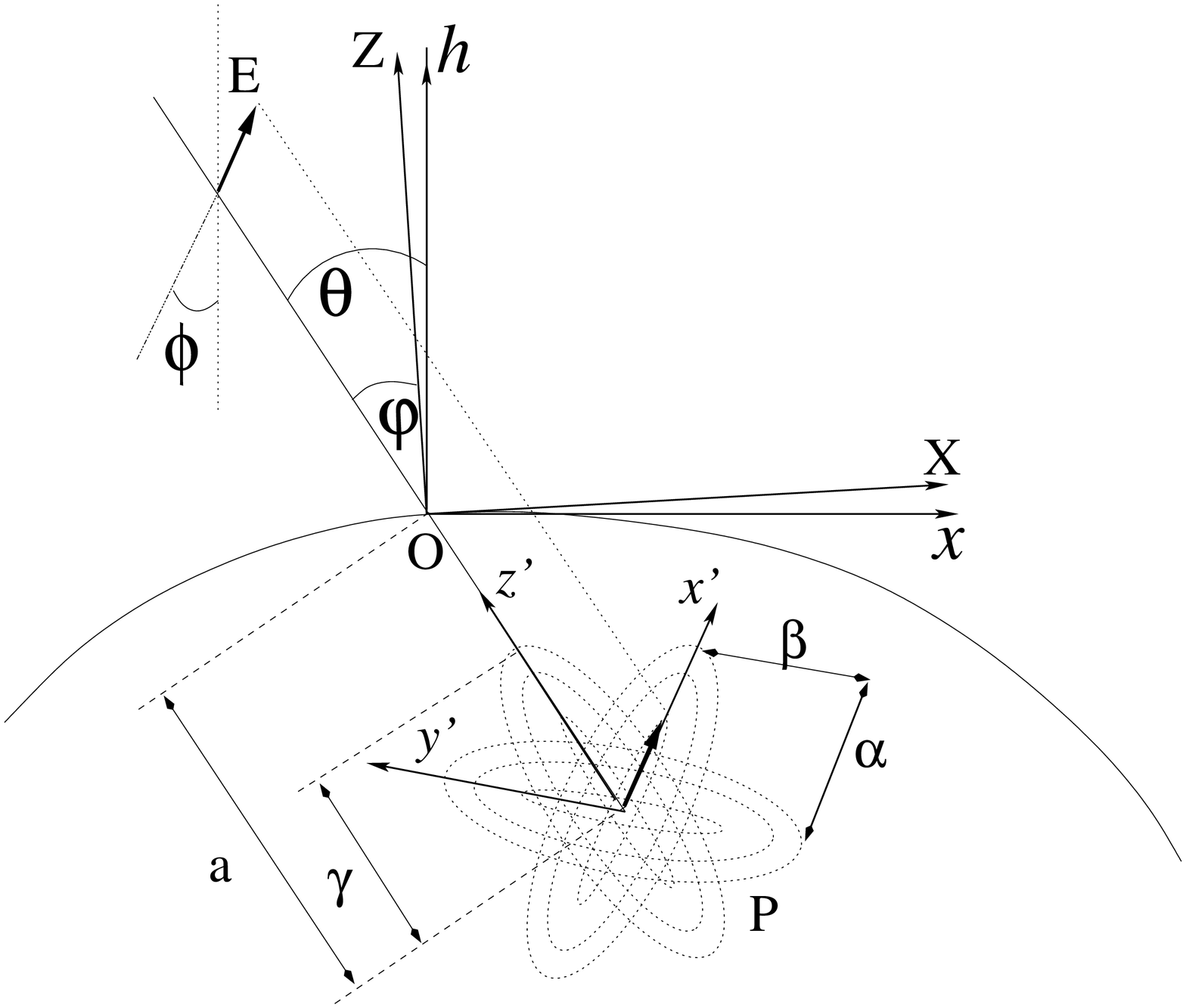}
\put(-70,160){\makebox(0,0){\includegraphics[width=3cm,clip=true]{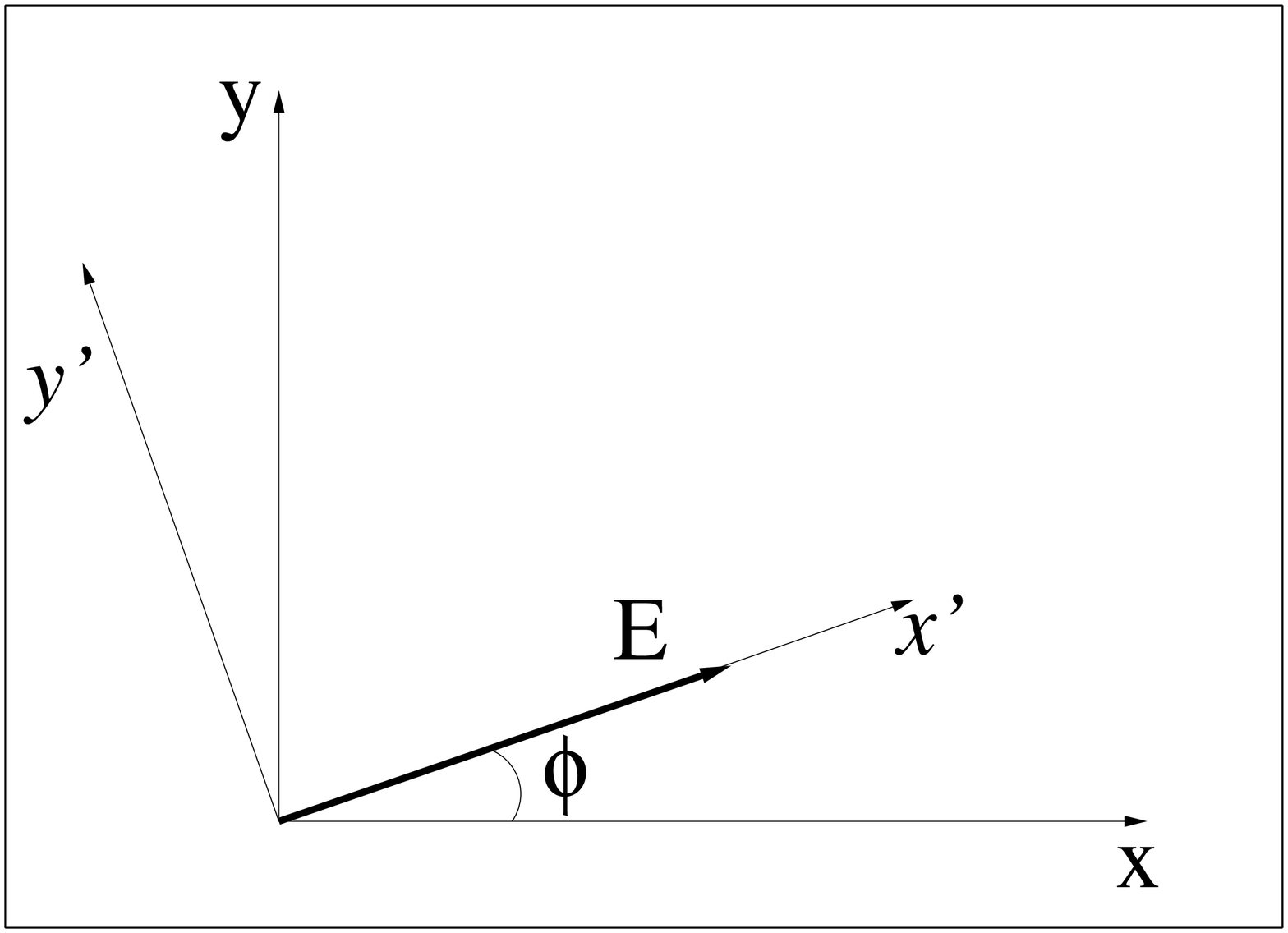}}}
\caption{Reference frames for the computation of the erosion velocity: 
(${\it x',y',z'}$) is reference frame of incoming beam where ${\it x'}$ axis is
parallel to the vector ${\bf E}$ of electromagnetic field;
($X,Y,Z$) correspond to the local coordinate frame, where $Z$ axis is 
parallel to the local normal to the surface, while (${\it x,y,h}$)
denotes the laboratory frame of coordinates with $h$ axis
perpendicular to the flat substrate.
The dotted ellipses are Gaussian distribution of deposited energy with halfwidth 
$\alpha, \beta, \gamma$ along ${\it x',y',z'}$ axis, respectively. 
The incidence angle measured in the local reference frame is $\varphi$, and $\theta$ 
in the laboratory frame. $\phi$ is the angle between the incidence plane ($X,Z$) 
and ${\bf E}$. Insert: view from top along $h$ for $\theta=0$.}
\end{figure}  
The local correction to the incident energy flux  is now given by
$\Psi({\bf r})=\xi I_0 (\cos \varphi + (\partial_X Z) \sin \varphi)$, 
where $\xi$ is the absorption coefficient of the material and
$I_0$ is the laser intensity. In order to describe the surface profile 
in the neighborhood of O we took into account cross-terms of the type
$\sim XY$
\begin{equation}
Z(X,Y)\simeq-\frac{1}{2}\left(\frac{X^2}{R_X}+\frac{Y^2}{R_Y}\right) -\frac{XY}{R_{XY}},
\label{taylor:ii}
\end{equation}
As in \citep{CUERNO} we assume that the radii of curvature $R_X,R_Y,R_{XY}$ of 
the surface are much larger than the penetration depth a \cite{SNOS}, so that only terms
up to first order in $a/R_X,a/R_Y$ and $a/R_{XY}$ are kept. The integration 
results in the erosion velocity $V(\varphi,\phi,R_X,R_Y,R_{XY})$ as a function
of angles $\varphi, \phi$ and the curvatures 
$1/R_X=-\partial^2 Z/\partial X^2$, $1/R_Y=-\partial^2 Z/\partial Y^2$ and
$1/R_{XY}=-\partial^2 Z/\partial X\partial Y$.
Next we perform expansions in powers of derivatives of h(x,y,t) and rewrite 
V in terms of the laboratory coordinates (x,y,h) \citep{CUERNO}.
\begin{equation}
\frac{\partial h(x, y, t)}{\partial t}=-v(\phi,\varphi, R_X, R_Y, R_{XY})\sqrt{1+(\nabla h)^2}
\label{lab:coord}
\end{equation}
We complete Eq.~\eqref{lab:coord} by adding the following physical processes: 
the surface self-diffusion effects $- K\nabla^2(\nabla^2 h)$ where 
$K$ is the relaxation rate due to thermally activated surface diffusion, 
together with the fluctuations $\eta (x,y,z)$
(short noise) in the flux of the bombarding particles. Finally, we obtain 
the equation of motion known as an anysotropic noisy Kuramoto-Sivashinsky 
\citep{KURAMOTO} equation where the coefficients are now the complex functions  of two angles $\phi$ and $\theta$.
In order to simplify our consideration and because of the reasons discussed below 
we write  here the equation for the case of the normal incident  ($\theta=0$) and 
in the reference frame rotated by the angle $\phi$ 
(see Insert Fig.1) that means that $\nu_{xy}=(\nu_x-\nu_y) \tan (2\phi)=0$ and 
$\lambda_{xy}=-(\lambda_x-\lambda_y) \tan (2\phi)=0$ 
\begin{eqnarray}
\dot{h}_t&=&
\nu_x h_{xx} + \nu_y h_{yy} + \frac{\lambda_x}{2} h^2_x + \frac{\lambda_y}{2} h^2_y \nonumber \\
&-& K\nabla^2(\nabla^2 h) + \eta (x,y,z)
\label{kuramot}
\end{eqnarray}
where the coefficients are given by 
\begin{eqnarray}
\nu_{x,y}&=&-\frac{F}{2}\sigma_{\gamma}\Omega_{x,y},\,\,\,
\lambda_{x,y}=-\frac{F}{\gamma}[1+(\sigma_{\gamma}^2-1)\Omega_{x,y}], \nonumber \\
\Omega_{x,y}&=&\Delta [1 \pm \Pi], \nonumber \\
\Delta&=&\frac{1}{2}(\frac{\sigma^2_\gamma}{\sigma^2_\alpha}+\frac{\sigma^2_\gamma}{\sigma^2_\beta}), 
\Pi=\frac{\sigma^2_{\beta}-\sigma^2_{\alpha}}{\sigma^2_{\beta}+\sigma^2_{\alpha}}   
\label{surfev}
\end{eqnarray}
and $F=\frac{I_0}{\sqrt{2\pi}} exp (-\frac{\sigma_{\gamma}^2}{2})$, $\sigma_{\iota}=\frac{a}{\iota}$  
$(\iota=\alpha,\beta,\gamma)$. Moreover,  we neglected here the erosion velocity $v_0$ which
does not affect the ripple characteristics, such as ripple wavelength and the ripple amplitude, and
can be eliminated by the transformation $\tilde{h}=h+v_0 t$. 

The basic role of the introduced above  parameters
$\Delta$ and $\Pi$ can be clarified
by means of the linear stability analysis of  equation ~\eqref{kuramot}.
\begin{figure}[tbh]
\includegraphics[width=8cm,clip=true]{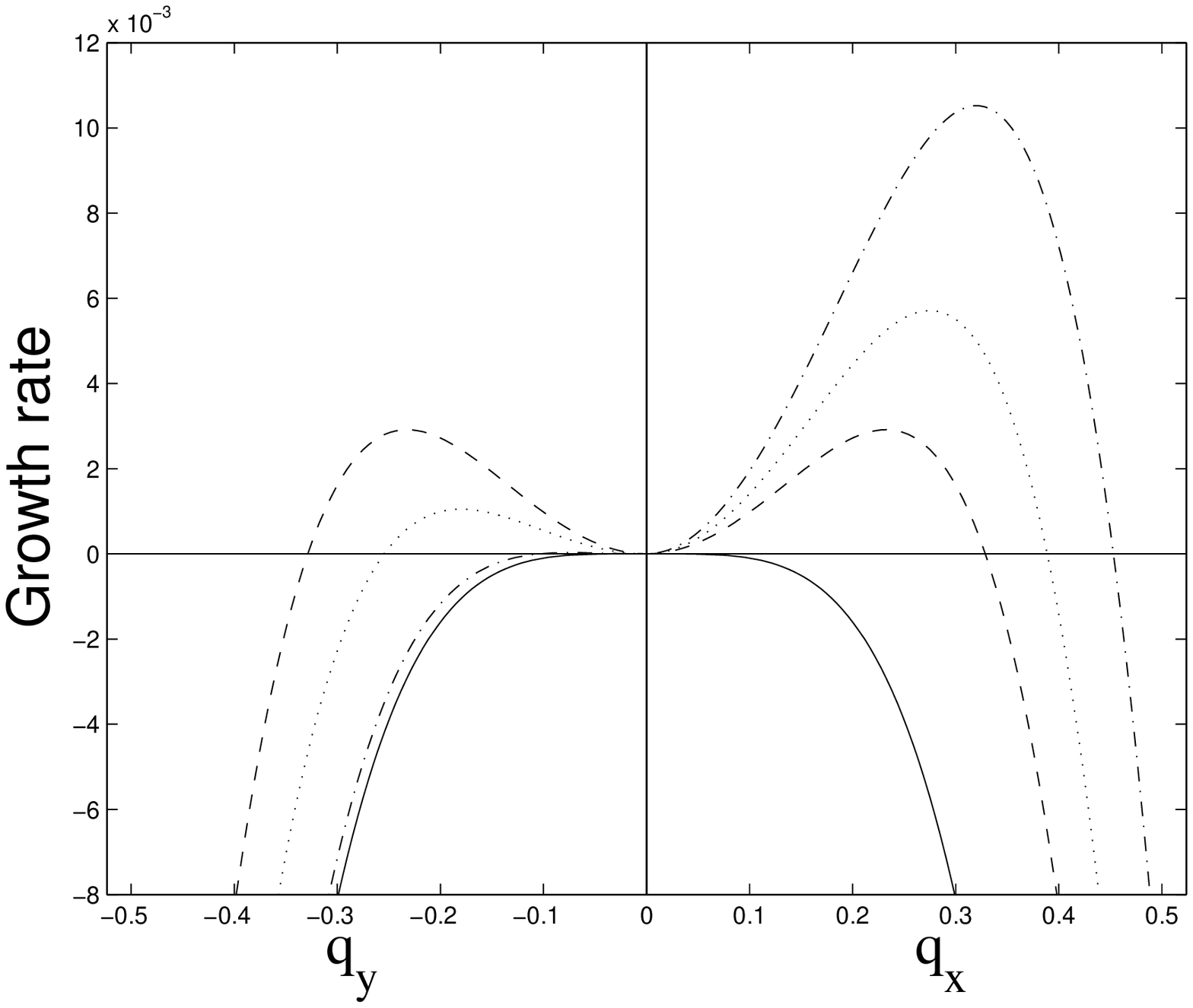}
\put(-165,148){\makebox(0,0){\includegraphics[width=2.5cm,clip=true]{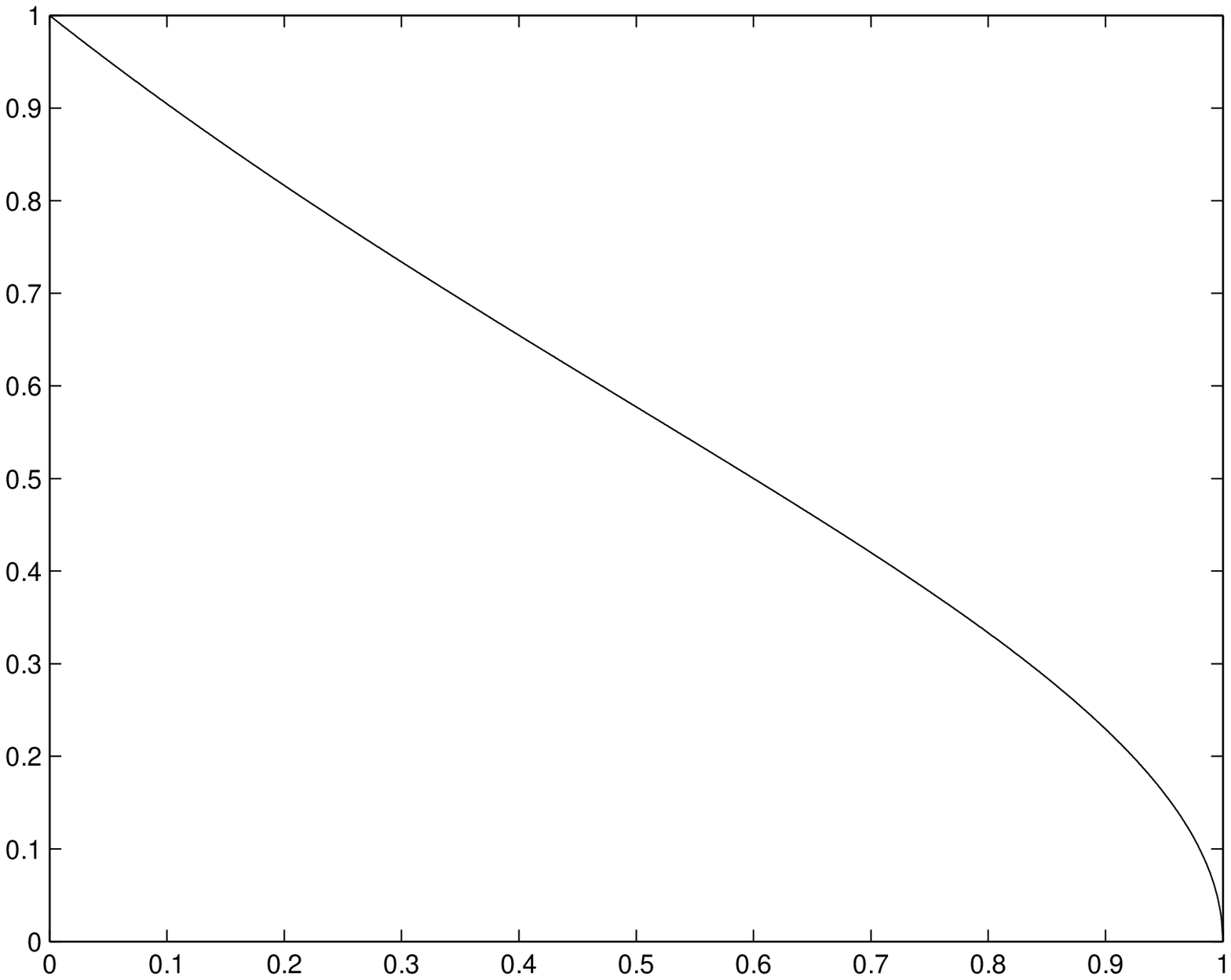}}}
\caption{Stability of Fourier modes for equation ~\eqref{kuramot} as a function
of the wave vector $q_x$ (right) and $q_y$ (left), for various values of the
parameters:$\Delta=0$,$\Pi=0$ (solid line); $\Delta=1$,$\Pi=0$ (dashed line);
$\Delta=1$,$\Pi=0.4$ (dotted line); $\Delta=1$,$\Pi=0.9$ (dash-dotted line).
Insert: $l_x/l_y$ as a function of $\Pi$.}
\end{figure} 
The Fig. 2 shows the linear growth rate along $q_x$ and $q_y$ for
various values of $\Delta$ and $\Pi$. For $\Delta=0$ the uniform
state is stable, whereas for nonzero $\Delta$ and $\Pi=0$ the instability
sets both along $q_x$ and $q_y$. An increase of $\Pi$ induces
the asymmetry in $q_{xy}$-plain thus for $\Pi \sim 1$ the 
instability along $q_y$ is suppressed. Thus, the deviation from
isotropic energy distribution breaks the rotation symmetry in $xy$ plain
and lids to an anysotropic linear instability.
For the sake of definiteness we will fix $\Delta=1/2$ that corresponds
to the condition $\alpha^2+\beta^2=\gamma^2$. 
From ~\eqref{kuramot} follows that the parameter $\Pi$ contributes  to the both tension coefficients
\begin{equation}
\nu_{x,y} = -\frac{F\sigma_\gamma}{4} (1 \pm \Pi),
\end{equation}
One can see that the surface tension coefficients are negative for the normal incidence
and in general are not equal to each other due to
the fact that the direction of the laser polarization breaks the symmetry along the surface. Consequently, we expect
the instability to the ripple formation with wavelength $l_i=2\pi\sqrt{2K/|\nu_i|}$, where i refers to the
direction (x or y) along which the associated $\nu_i$ ($\nu_x$ or $\nu_y$) is largest.
Thus, in our case of polarization along x-axe  for $\nu_x<\nu_y<0$  which holds when $0<\Pi<1$
the ripple structure is oriented in the x direction. 
Moreover, the ripple wavelength has now the following dependence from $\Pi$
\begin{equation}
l_{x,y}=4\pi \sqrt{\frac{2K}{F\sigma_\gamma(1 \pm \Pi)}}
\end{equation}
Thus, for $\Pi=0$ the wavelength $l_x=l_y$. 
The increase of $\Pi$ slightly reduces $l_x$ whereas $l_y$ become to be
very large for $\Pi \to 1$ (linear polarization along x axis). 
The inset of Fig.2 shows the relation $l_x/l_y=\sqrt{(1-\Pi)/(1+\Pi)}$. 
Thus, the parameter $\Pi$ and the angle $\phi$ allow us to change the orientation and topology of ripples
that means that the parameter $\Pi$ is
a control parameter which we will call in the following the degree of polarization.
We will consider three cases:
(i) circular polarization, $\Pi=0$ ($\sigma_\alpha=\sigma_\beta$);
(ii) linear laser polarization, $\Pi=1$ ($\sigma_\alpha<<\sigma_\beta$),
and (iii) elliptic polarization, $0<\Pi<1$ ($\sigma_\alpha<\sigma_\beta$).

Another quantity which can change the orientation of ripple is  incidence angle $\theta$.
Thus, for angles $\theta$ less than a critical angle $\theta_c$,
the wave vector of the modulations is parallel to the component of the beam
in the surface plane \cite{BRADLEY}.
\begin{figure}[tbh]
\includegraphics[width=8cm,clip=true]{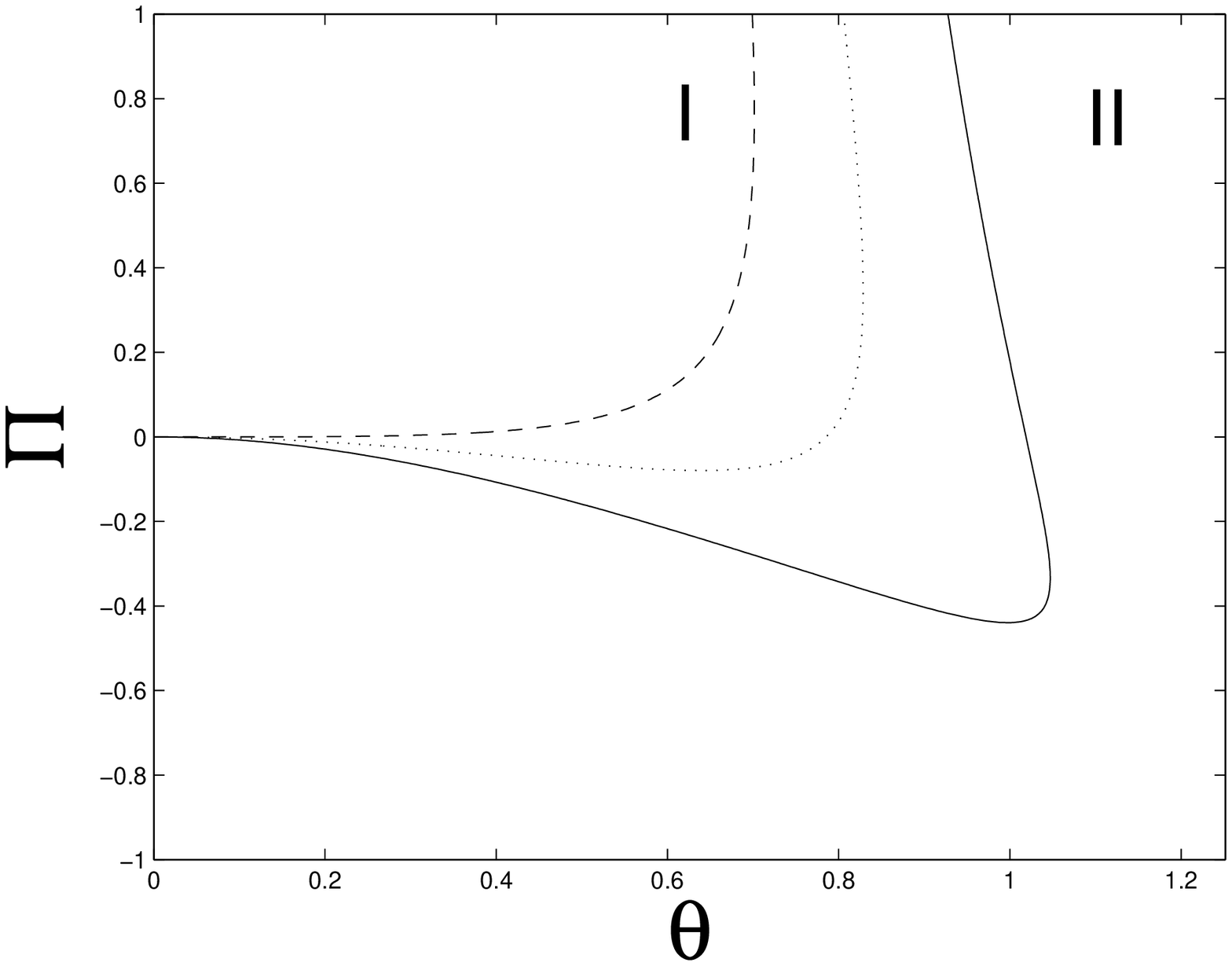}
\put(-165,50){\makebox(0,0){\includegraphics[width=1.5cm,clip=true]{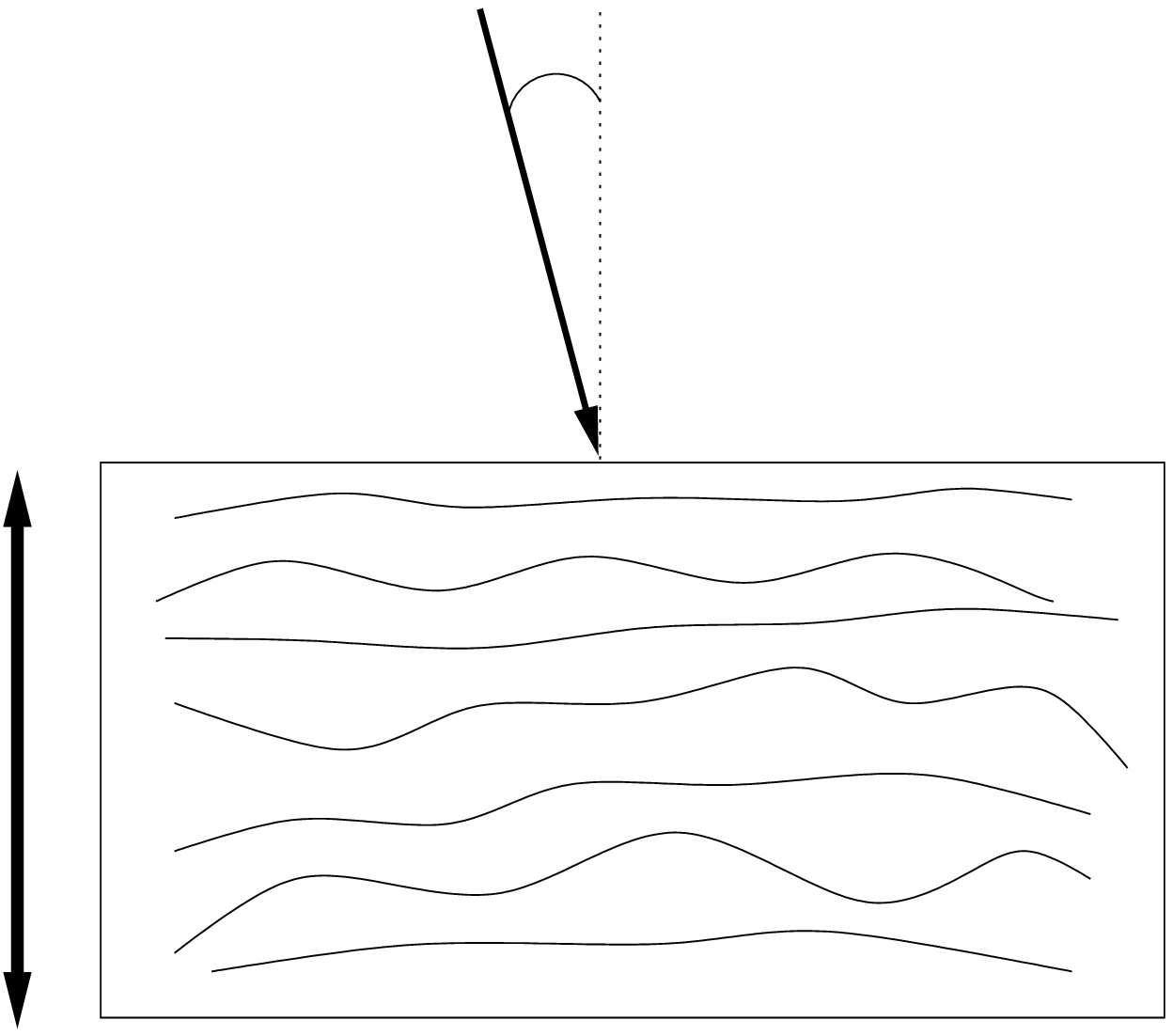}}}
\put(-165,150){\makebox(0,0){\includegraphics[width=1.5cm,clip=true]{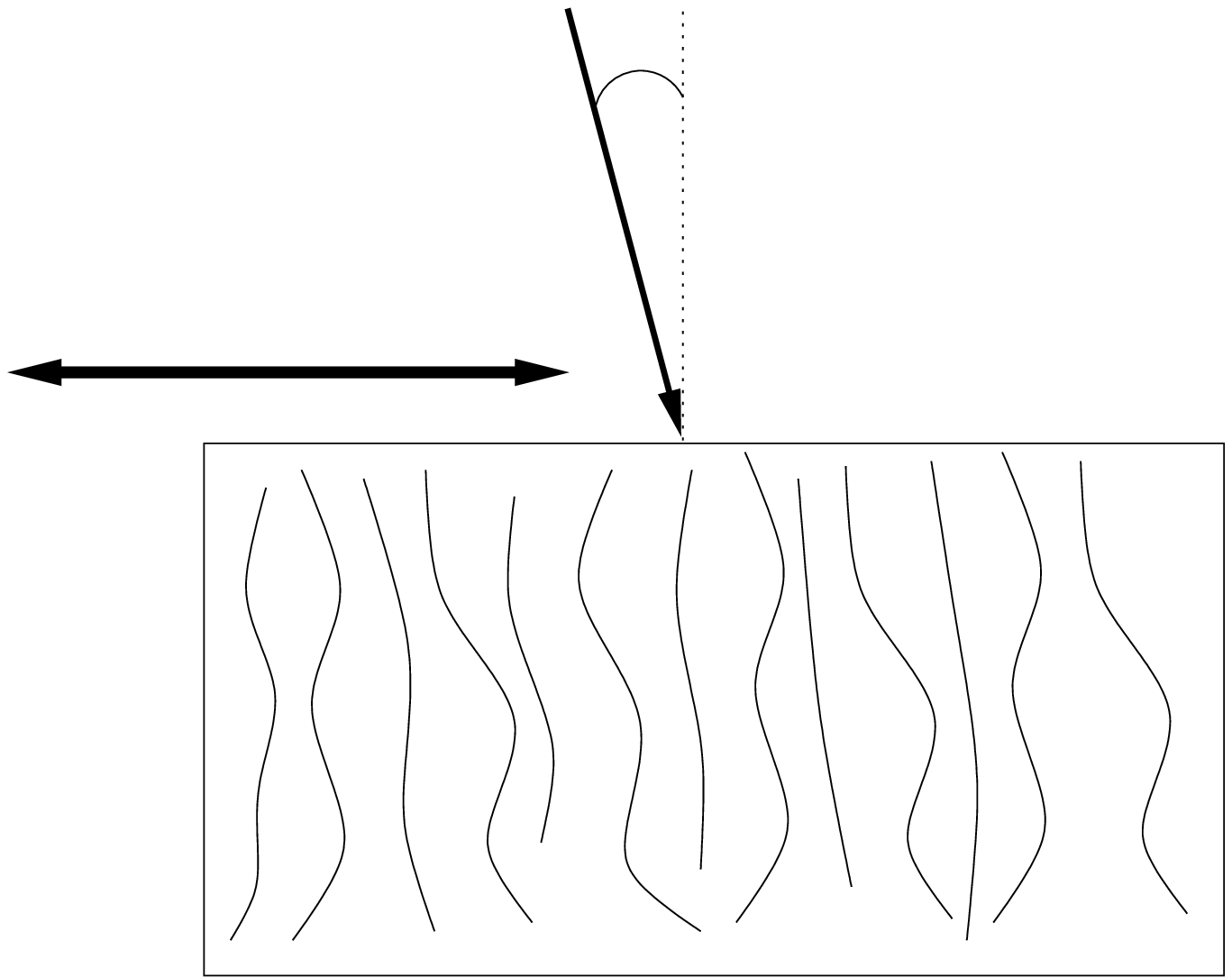}}}
\put(-31,45){\makebox(0,0){\includegraphics[width=1.5cm,clip=true]{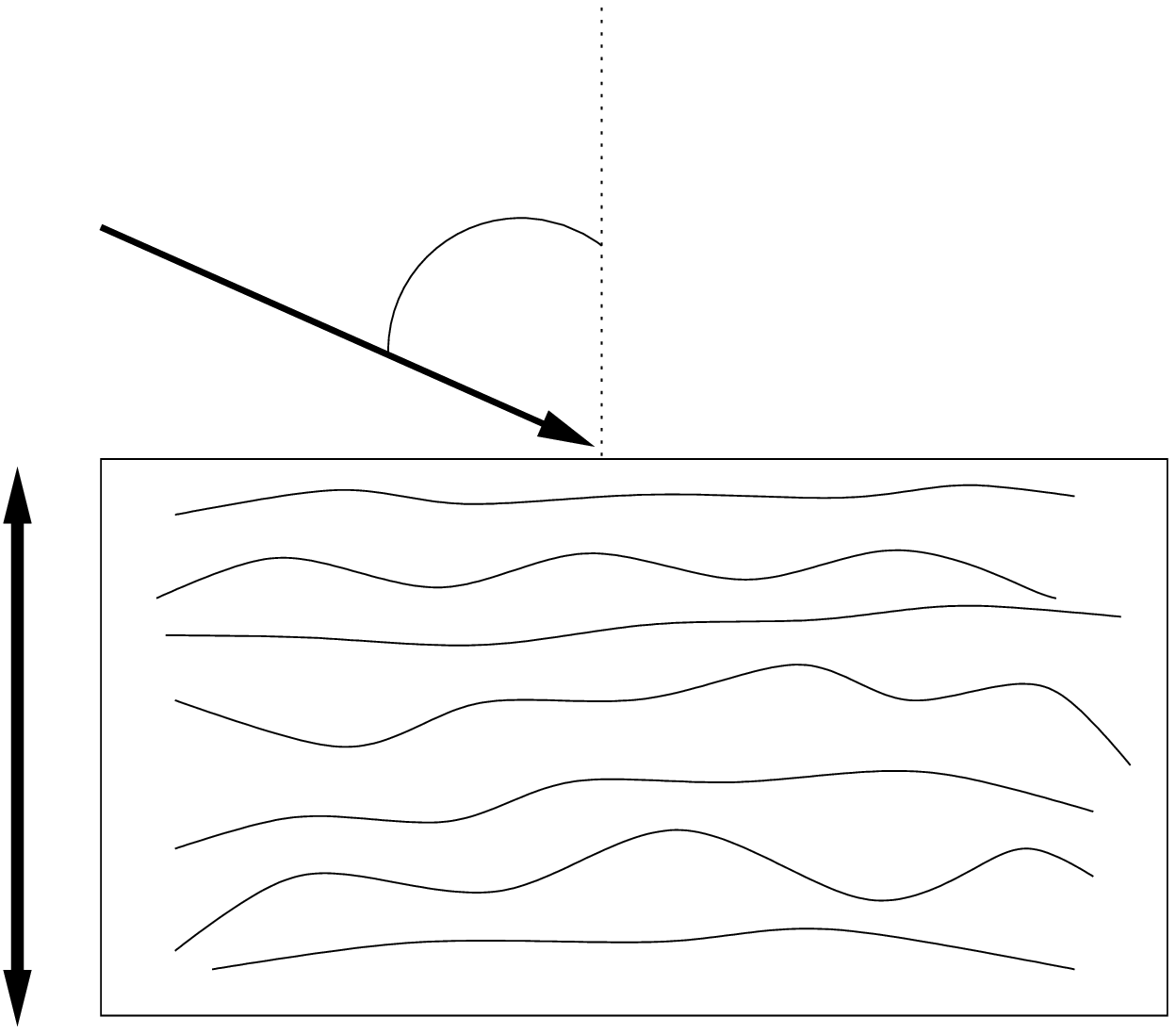}}}
\caption{Ripple orientation phase diagram for two values of the
parameter $\sigma_\gamma=1,1.5,2$ (dashed, dotted, solid line,respectively).
Region I: $\nu_x<\nu_y$; Region II: $\nu_x>\nu_y$. Inserts show the incidence angle,
the direction of ripples and the vector E.}
\end{figure}
By this means the orientation of ripple due to polarization and due to incoming angle
can compete or work together. 
Fig. 3 displays the ripple orientation phase diagram ($\theta,\Pi$) for three
values of parameter $\sigma_\gamma$ (negative values of $\Pi$ correspond to the vector E along 
y axis). 
The boundary is defined by $\nu_x(\Pi,\theta)=\nu_y(\Pi,\theta)$ and
separates the region I ($\nu_x<\nu_y$) and region II ($\nu_x>\nu_y$).
In the case of $\Pi=0$ (circular polarization) and the normal incidence of laser beam
the dot  structures  are expected in our model.
The increase of $\Pi$ for small $\theta$ aligned the ripples along the vector $E$.
Whereas the growth of $\theta$ can still change the direction of ripple for $\Pi>0$
at some $\theta_c$, which depends from $\sigma_\gamma$, the laser polarization 
completely suppress the transition for negative $\Pi$. The interesting result,
which can be tested experimentally,
is the possible reorientation of ripple in the case of weak polarization $-0.4<\Pi<0$ 
for $\sigma_\gamma=2$. In any case, one have to perform detailed experimental investigation
the angle dependence for various values of $\Pi$  in order to fix the parameter $\sigma_\gamma$.
Moreover, our model allows us to understand the nature of LIPPS which are oriented parallel 
to the laser p-polarization at a relatively large incident angle. 
Nevertherless, we have to note here that our approximation 
may be not valid for the large angles of the incidence $\theta$
where the effects of interaction of laser radiation with the material surface can not be
neglected. Therefore, we limit ourselves in the following to the presentation of our results only for the normal
incidence of laser beam and set $\sigma_\gamma=2$.
The dependence of ripples structures on the angle of the incidence for large $\theta$  will be
discussed somewhere else.
\begin{figure}[tbh]
\includegraphics[width=8cm,clip=true]{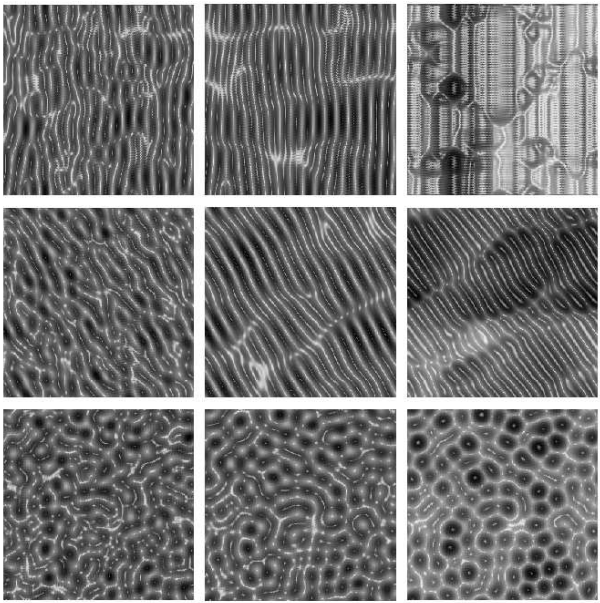}
\caption{Grey-scale plot of a surface, showing the surface morphologies for (from top to bottom):
$\phi=0$,$\Pi=1$;  
$\phi=30^o$,$\Pi=0.5$; 
$\phi=45$,$\Pi=0$,
and for various values  (from left to right) of time: $t<<t_c$; $t=t_c/2$ and $t \approx t_c$}
\end{figure}

Having analyzed the linear regime we now precede by investigating the influence 
of nonlinear terms $\lambda_x$ and $\lambda_y$. 
As shown early \citep{PARK}, there is a clear separation of the linear and the nonlinear behaviour in time
in such way that in the linear regime up to a crossover time $t_c$ the nonlinear terms as would be absent
whereas the nonlinear terms take over after $t_c\sim (K/\nu^2)ln (\nu/\lambda)$ and completely determine
the surface morphology which depends on the relative signs of $\lambda_x$ and $\lambda_y$. 
For our choice $\sigma_\gamma=2$ both $\lambda_x$ and $\lambda_y$ are negative and depend from $\Pi$, that means 
the disappearance of ripples above $t_c \sim t_c(\Pi=0)/(1+\Pi)^2$, which is decreased with increasing of $\Pi$  
and the appearance of kinetic roughening. 

The surface morphologies obtained
by numerically integrating the equation ~\eqref{kuramot}  for various values of $\phi$, $\Pi$ and time
show in Fig.4 that the anisotropy in the energy deposition can describe 
the correlation of ripple orientation with laser polarization and
saturation of ripples with increasing of time. 
However, in spite of remarkable concurrence  the calculated and the  
ablated surfaces \cite{VARLAMOVA}, a direct comparison with experimental date is  difficult
because the relevant mechanism which is responsible for the energy anisotropy is still under discussion. 
Whereas in metalls and semiconductors the incident laser radiation can excite the  plasma wave in the electron plasma 
produced by multiphoton ionization \cite{SHIMOTSUMA}, the electron density in dielectrics is too low for the
intensities near the damage threshold. Instead, it was proposed, that 
the femtosecond laser pulses could induce high inhomogeneous
ionization under laser radiation inside the transparent solids and the dielectrics,
forming nano-droplets of plasma by means of the "forest-fire"
multiphoton and avalanche ionization \cite{GAIER}. 
Interaction of infrared laser radiation with
nano-plasma-dielectric composites can excite the surface plasmons polaritons (SPP)
in the nano-particles and induce the giant enhancement of local electric fields.
In contrast to \cite{BHARDWAJ}, where the underdense nanoplasmas  grows
into nanosheets orientated with their normal parallel
to the laser polarization, we consider the energy deposition along  the laser polarization
due to the powerflow distribution an oblate nanoparticle \cite{BASHEVOY}.
Thus, the concentration of the powerflow along the vector ${\bf E}$
of electromagnetic field near the nano-particles leads to the local instability of the lattice
that could bring a collapse of the atomic structure about.
We think that the surface atoms of such annealed  areas have a good chance to be
sputtered at the time scale of electron-phonon relaxation.

In summary, we have shown that the correlation of ripples orientation 
with laser polarization can be consistently described within a model 
where the  polarization induces an anisotropy in the energy distribution
and causes the symmetry break on the surface. The model accounts 
for experimental features of laser induced surface modulations and leads
to the numerous predictions which can be directly tested experimentally.
Moreover, our results support the nonlinear self-organized mechanism of formation
the ripples on the surface of solids.
However, a more detailed investigation of the effective mechanism which is responsible for 
the  anisotropy of energy transfer  remains an interesting issue for future work.




\begin{thebibliography}{9}
\bibitem{SNOS} This penetration depth is differ from the real penetration depth and means 
the distance from the surface to the bulk, at which  the scattering of light in nano-plasmas have 
influence on the surface atoms.
\end{thebibliography}
\end{document}